\documentclass[%
 reprint,
superscriptaddress,
 amsmath,amssymb,
 aps,
prl,
]{revtex4-2}

\usepackage{graphicx}
\usepackage{dcolumn}
\usepackage{bm}

\usepackage{xcolor}

\begin{document}


\title{Magnetic Control of Soft Chiral Phonons in PbTe}

\author{Andrey Baydin}
\thanks{These authors contributed equally to this work}
\affiliation{Department of Electrical and Computer Engineering, Rice University, Houston, Texas 77005, USA}%
\affiliation{Smalley-Curl Institute, Rice University, Houston, Texas, 77005, USA}
\email{baydin@rice.edu}
    
\author{Felix G.\ G.\ Hernandez}
\thanks{These authors contributed equally to this work}
\affiliation{Instituto de F\'{i}sica, Universidade de S\~{a}o Paulo, S\~{a}o Paulo, SP 05508-090, Brazil}
\email{felixggh@if.usp.br}

\author{Martin Rodriguez-Vega}
\thanks{These authors contributed equally to this work}
\affiliation{Theoretical Division, Los Alamos National Laboratory, Los Alamos, New Mexico 87545, USA \looseness=-1}

\author{Anderson K.\ Okazaki}
\affiliation{Laboratório Associado de Sensores e Materiais, Instituto Nacional de Pesquisas Espaciais, São José dos Campos, SP 12201-970, Brazil}

\author{Fuyang Tay}
\affiliation{Department of Electrical and Computer Engineering, Rice University, Houston, Texas 77005, USA}
\affiliation{Applied Physics Graduate Program, Smalley-Curl Institute, Rice University, Houston, Texas, 77005, USA}

\author{G.\ Timothy Noe II}
\affiliation{Department of Electrical and Computer Engineering, Rice University, Houston, Texas 77005, USA}%

\author{Ikufumi Katayama}
\affiliation{Department of Physics, Graduate School of Engineering Science, Yokohama National University, Yokohama 240-8501, Japan}

\author{Jun Takeda}
\affiliation{Department of Physics, Graduate School of Engineering Science, Yokohama National University, Yokohama 240-8501, Japan}

\author{Hiroyuki Nojiri}
\affiliation{Institute for Materials Research, Tohoku University, Sendai 980-8577, Japan}

\author{Paulo H.\ O.\ Rappl}
\affiliation{Laboratório Associado de Sensores e Materiais, Instituto Nacional de Pesquisas Espaciais, São José dos Campos, SP 12201-970, Brazil}

\author{Eduardo Abramof}
\affiliation{Laboratório Associado de Sensores e Materiais, Instituto Nacional de Pesquisas Espaciais, São José dos Campos, SP 12201-970, Brazil}

\author{Gregory A.\ Fiete}
\affiliation{Department of Physics, Northeastern University, Boston, Massachusetts 02115, USA}
\affiliation{Department of Physics, Massachusetts Institute of Technology, Cambridge, Massachusetts 02139, USA}

\author{Junichiro Kono}
\email{kono@rice.edu}
\affiliation{Department of Electrical and Computer Engineering, Rice University, Houston, Texas 77005, USA}%
\affiliation{Smalley-Curl Institute, Rice University, Houston, Texas, 77005, USA}
\affiliation{Department of Physics and Astronomy, Rice University, Houston, Texas 77005, USA}%
\affiliation{Department of Material Science and NanoEngineering, Rice University, Houston, Texas 77005, USA}

\date{\today}

\begin{abstract}
PbTe crystals have a soft transverse optical phonon mode in the terahertz frequency range, which is known to efficiently decay into heat-carrying acoustic phonons, resulting in anomalously low thermal conductivity. Here, we studied this phonon via polarization-dependent terahertz spectroscopy. We observed softening of this mode with decreasing temperature, indicative of incipient ferroelectricity, which we explain through a model including strong anharmonicity with a quartic displacement term. In magnetic fields up to 25\,T, the phonon mode splits into two modes with opposite handedness, exhibiting circular dichroism.  Their frequencies display Zeeman splitting together with an overall diamagnetic shift with increasing magnetic field. Using a group-theoretical approach, we demonstrate that these observations are the result of magnetic field-induced morphic changes in the crystal symmetries through the Lorentz force exerted on the lattice ions. Thus, our study reveals a novel process of controlling phonon properties in a soft ionic lattice by a strong magnetic field. 

\end{abstract}

\maketitle

Phonons, collective excitations of a crystal lattice, determine many physical properties of solids, including electrical, thermal, and optical. By exciting and controlling particular phonon modes, one can modify material properties or even induce quantum effects such as superconductivity. Usually, phonons are insensitive to magnetic fields. Unlike spectral features related to electrons, which respond to magnetic fields through their orbital and spin magnetic moments, phonon-related features typically remain independent of magnetic fields. While some phonons are predicted to carry an orbital magnetic moment due to ionic cyclotron motion~\cite{JuraschekEtAl2017PRM,JuraschekSpaldin2019PRM}, a phonon Zeeman effect has been experimentally observed  either via strong electron-phonon coupling~\cite{ChengetAl2020NL} or in strong paramagnets~\cite{Schaack_1976}. 

Recently, there has been much interest in phononic properties of lead telluride (PbTe) to throw light on the microscopic origin of its exceptional thermoelectric properties~\cite{PhysRevLett.112.175501,NatComm7.12291,PhysRevB.97.184305,PhysRevX.10.041029}. Reports have related the low thermal conductivity to anomalous lattice dynamics with giant anharmonic coupling of the longitudinal acoustic (LA) modes with the transverse optical (TO) phonon mode~\cite{DelaireetAl2011NM,XiaetAl2018APL,LuetAl2018PRB,Ribeiro2018}. Softening of this TO phonon mode with decreasing temperature has been observed and interpreted as an onset of ferroelectric instability, as seen in other good thermoelectric materials; their phase transition temperatures depend on the carrier concentration, indicating the importance of electron-phonon coupling~\cite{NatComm7.12291,PhysRevResearch.2.012048,PhysRevB.102.024112,PhysRevB.102.115204}. These results for the anharmonicity and incipient ferroelectricity of PbTe have been obtained in a small number of experimental studies on reflectivity spectroscopy~\cite{Burkhard:77} and, more recently, inelastic x-ray scattering~\cite{NatComm7.12291} and neutron scattering measurements~\cite{DelaireetAl2011NM}.

The ionic nature of the PbTe lattice is expected to lead to even richer lattice dynamics in the presence of a magnetic field, as it may host circularly polarized phonons carrying a magnetic moment~\cite{JuraschekSpaldin2019PRM}. Such magnetic moments arise from the orbital angular momentum of ions, due to the circular motion of ions with different masses, and can be observed as an energy splitting between left-hand and right-hand phonon polarization in external magnetic fields~\cite{Schaack_1976}. Furthermore, a magnetic field can produce a Lorentz force on the ions in the lattice, which induces morphic changes in the crystal~\cite{ANASTASSAKIS1971563,ANASTASSAKIS19721091}. However, effects of strong magnetic fields on these anharmonic phonons have not been investigated.

In this Letter, we report results of polarization-dependent terahertz time-domain spectroscopy (THz-TDS) studies of single-crystalline PbTe films (1.6 $\mu$m thick) epitaxially grown on (111) BaF$_2$ substrates in high magnetic fields. The temperature dependence of the phonon conductivity displayed a behavior consistent with quartic anharmonicity that modifies the frequency and line shape. The optical conductivity in intense magnetic fields up to 25~T revealed unique magnetophononic effects -- phononic magnetic circular dichroism (MCD), a phonon Zeeman splitting, and a phonon diamagnetic shift. The observed Zeeman splitting was 3 orders of magnitude larger than what has been predicted for PbTe using density functional theory based on harmonic phonons, indicating the importance of anharmonicity in the observed magnetic field induced phenomena. 

\begin{figure}
    \centering
    \includegraphics{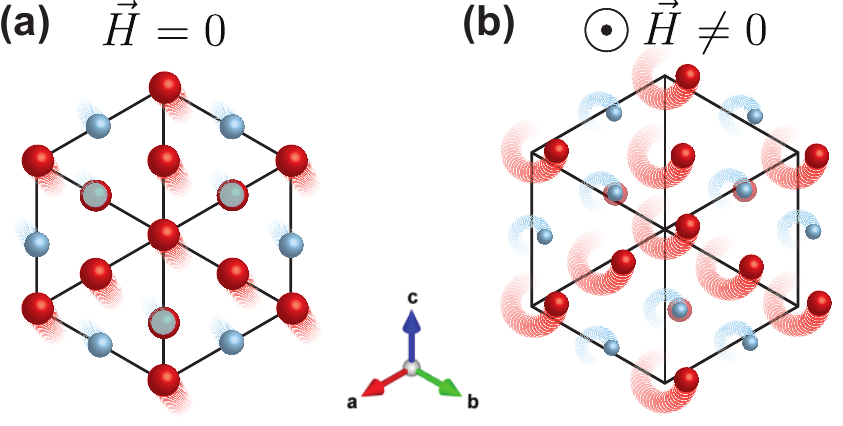}
    \caption{Schematic of the PbTe crystal structure with real-space TO lattice displacements (a)~without and (b)~with a magnetic field applied perpendicular to the lattice plane. Blue (red) spheres represent Te (Pb) ions.}
    \label{fig:lattice_phonon}
\end{figure}

PbTe belongs to the space group $Fm\bar 3 m$ with $O_h$ point group. The Pb and Te atoms form an arrangement of interpenetrating face-centered cubic lattices~\cite{Xu2012}.  Our group theory analysis, performed using the Bilbao Crystallographic server~\cite{bilbao-sam}, indicates that the irreducible representation of the optical mode is $T_{1u}$, which is infrared active and triply degenerate without considering the TO-LO splitting. Within harmonic first-principles calculations (see Supplementary Material~\cite{supplemental} for details), we estimate the phonon frequencies to be $\approx 1.25$~THz for the doubly degenerate TO phonons and $\approx 3.32$~THz for the LO phonon. The anharmonic coupling with acoustic phonons further changes this frequency~\cite{Ribeiro2018}. An incident light beam propagating in the $[111]$ direction couples linearly to these TO phonon modes, inducing ionic oscillations in a perpendicular plane. Anharmonic interactions can modify the details of this real-space picture, but we expect it to hold to a good approximation~\cite{Ribeiro2018}. Figure~\ref{fig:lattice_phonon} schematically shows the TO lattice displacement (a)~without and (b)~with a magnetic field.

\begin{figure}
    \centering
    \includegraphics{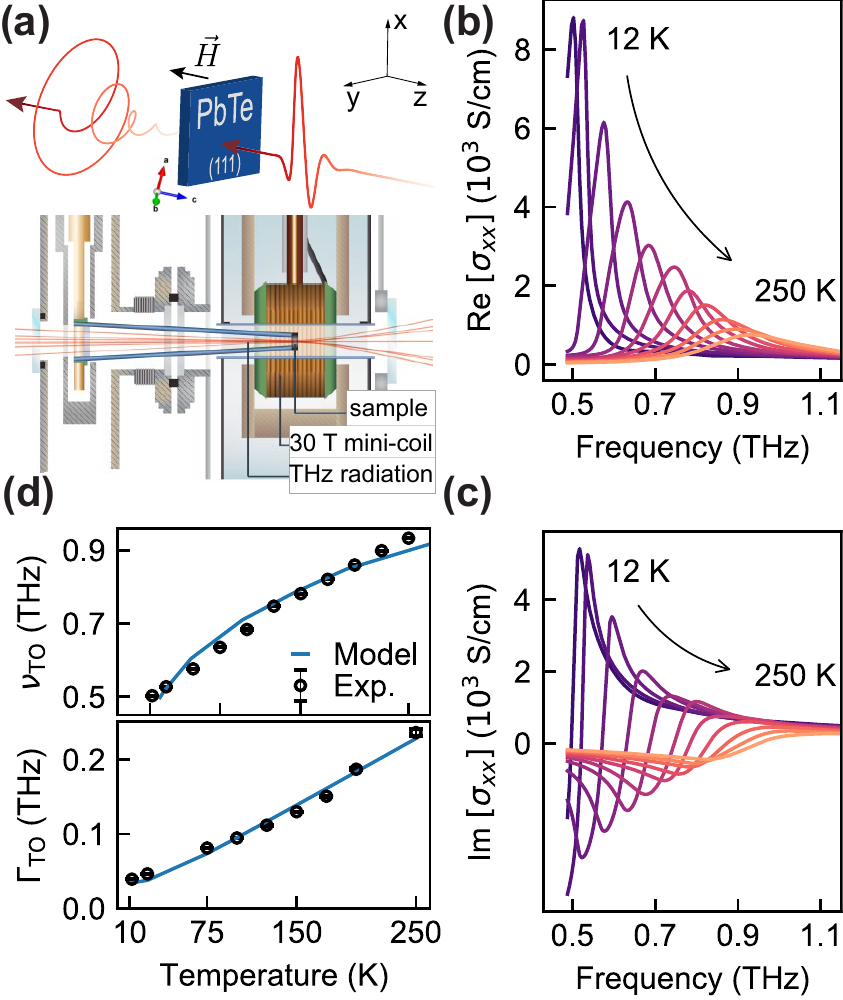}
    \caption{(a)~Schematic diagram for the THz time-domain spectroscopy in a magnetic field in transmission geometry [top] and RAMBO setup~\cite{NoeetAl2016OE} [bottom]. (b)~Real and (c)~imaginary parts of the conductivity of PbTe as a function of temperature without an applied magnetic field. (d)~TO phonon frequency [top] and linewidth [bottom] as a function of temperature from the experiments (circles) and the model (solid line) as described in the text.}
    \label{fig:temperature}
\end{figure}

We performed experiments by combining high magnetic fields and low temperatures with ultrashort laser pulses, which required advanced instrumentation that has been described previously~\cite{Baydinetal2021FO}. We show a schematic of our THz-TDS setup in Fig.~\ref{fig:temperature}(a) for transmission geometry (top) in continuous (Oxford Spectromag not shown) and pulsed (bottom) magnetic fields. In the first case, we used the output from a Ti:sapphire regenerative amplifier (1\,kHz, 150\,fs, 775\,nm, Clark-MXR, inc.)\ to generate THz radiation via optical rectification in a ZnTe crystal, and we probed it through electro-optic sampling in another ZnTe crystal. The sample was placed in a 9-T superconducting magnet (Oxford Instruments). In the second configuration, another Clark-MXR amplifier was used to pump a LiNbO$_3$ crystal for single-shot detection of THz radiation in magnetic fields up to 30\,T (RAMBO system)~\cite{NoeetAl2016OE}.

Figure~\ref{fig:temperature} shows the (b)~real and (c)~imaginary parts of the optical conductivity at different temperatures, which were extracted from the measured transmittance using conventional THz-TDS methods~\cite{NeuSchmuttenmaer2018JAP}. Both the real and imaginary parts of the conductivity indicate a strongly temperature-dependent resonance, which can be attributed to the TO phonon in PbTe. Figure~\ref{fig:temperature}(d) shows the frequency [top] and linewidth [bottom] of this TO phonon as a function of temperature obtained using Fano line shape fits~\cite{ChengetAl2020NL}. The black dots are from the experimental data. 

Interestingly, the TO phonon's linewidth temperature dependence is well fitted by the model for a harmonic optical phonon decaying into two acoustical phonons with finite momentum, as derived in Ref.~\cite{Klemens1966PR} (blue line in Fig.~\ref{fig:temperature}(d)[bottom]). Further experiments with improved frequency resolution could shed light on the contribution of the anharmonicity to the intrinsic linewidth (which could arise from the optical phonon interacting with the electron continuum or acoustic phonons) and its temperature dependence.

The TO phonon softens as the temperature decreases (see Fig.~\ref{fig:temperature}(d)[top]). This sublinear temperature dependence has been understood by strong anharmonic coupling between TO phonons and LA phonons with first principles calculations beyond the harmonic approximation~\cite{JochymEtAl2020SR, Ribeiro2018}. In this work, we take a group theory approach, which complements the first-principles studies in the literature and provides a unified perspective of the magnetic field effect, as we will discuss later. Since the phonons under investigation are infrared active (odd under inversion), the simplest, lowest-order anharmonic Hamiltonian one can construct to model our THz spectroscopy measurements is
\begin{equation}
   \mathcal H = \mathcal H_{\text{ph}} + Z^* X E_0,
\end{equation} 
where $\mathcal H_{\text{ph}}$ is the phonon Hamiltonian including a quartic anharmonic term~\cite{PhysRevB.79.214306}, $E_0$ is the incident THz electric field, $X$ is the infrared (TO) phonon real-space coordinate, and $Z^*=Z^*(H)$ is the phonon Born effective charge, which can depend on the applied magnetic field $H$, depending on the symmetries of the system. The phonon Hamiltonian can be written as~\cite{PhysRevB.79.214306},
\begin{equation}
    \mathcal H_{\text{ph}} =\frac{P^{2}}{2 m}+\frac{1}{2} k X^{2}+\frac{1}{4} \lambda X^{4}, 
\end{equation}
where $k$ is the spring constant, $m$ is the effective mass, $P$ is the momentum, and $\lambda$ is the anharmonic strength. We calculated the complex conductivity $\sigma(\omega)$ for this model, as detailed in the Supplemental Material~\cite{supplemental}. In the harmonic limit ($\lambda \rightarrow 0$), the model predicts no frequency temperature dependence. For finite $\lambda$, the phonon frequency increases as the temperature increases. In particular, in the strong anharmonicity regime, $\tilde \lambda \gg \tilde k$, where $\tilde{\lambda}={\hbar \lambda}/{m^{2} \omega_{0}^{3}}$ and $\tilde{k}={k}/{m \omega_{0}^{2}}$, and for a constant lifetime, theory and experimental data are in good agreement, as shown in Fig.~\ref{fig:temperature}(d). For a model with a temperature-dependent phonon lifetime and, the general conclusion remains valid, although some details hinge on the assumptions for the lifetime, as discussed in the Supplemental Material~\cite{supplemental}.

\begin{figure}
    \centering
    \includegraphics{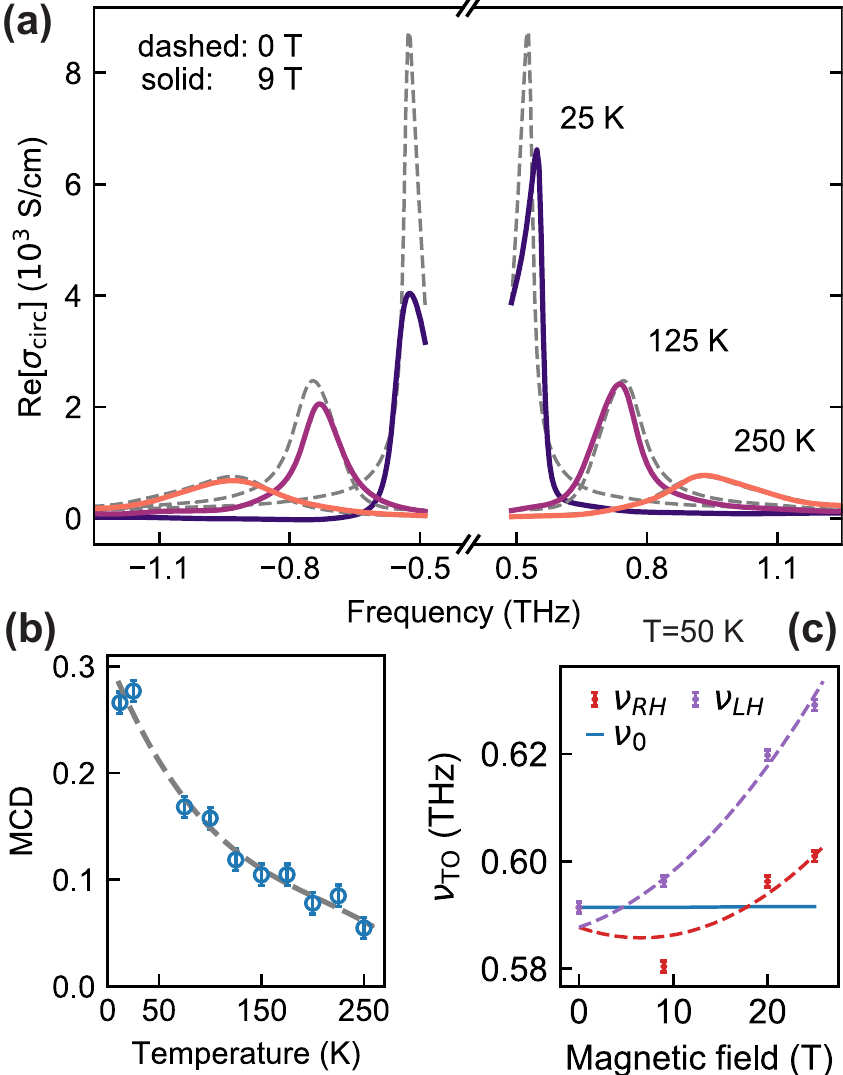}
    \caption{(a)~Optical conductivity at 0\,T (dashed line) and 9\,T (colored lines) for left-handed (LH) and right-handed (RH) phonons. (b)~Magnetic circular dichroism at 9\,T as a function of temperature. The dashed line is a guide to the eye. (c) Magnetic field induced frequency shift of the TO phonon at 50~K. The dashed lines are fits to the data using Eq.~\eqref{eq:fit}.}
    \label{fig:magnetic}
\end{figure}

In the presence of magnetic fields, we measured transmittance $T_{xx}$ and $T_{xy}$ with the detection polarizer along the $x$ and $y$ directions, respectively, while the input polarizer was set in the $x$ direction. The transmittance was then converted to a circular basis as $T_{R,L} = T_{xx} \pm iT_{xy}$, which was in turn converted into the conductivity in the circular basis~\cite{ChengetAl2019PRL}. 
Figure~\ref{fig:magnetic}(a) shows the real part of the conductivity in the circular basis for 0~T (dashed lines) and 9~T (colored lines) at different temperatures. The conductivity in the circular basis is depicted using positive and negative frequencies as the response to right-hand ($R$) and left-hand ($L$) polarized light, respectively. 

At high temperatures, the 0~T and 9~T data overlap, while at low temperatures, the TO phonon response in a magnetic field is different from that with no magnetic field. Moreover, the response to right-hand and left-hand polarized light varies, which indicates MCD of the TO phonon. We plot MCD = $(\sigma_R-\sigma_L)/(\sigma_R+\sigma_L)$ for different temperatures in Fig.~\ref{fig:magnetic}(b). We used the area under the phonon peak for $\sigma_{R,L}$. Figure~\ref{fig:magnetic}(c) shows the RH and LH TO phonon frequency as a function of applied magnetic field, together with fits (dashed lines), as described below. 

Recently, Cheng \textit{et al}.~\cite{ChengetAl2020NL} observed a large effective phonon magnetic moment in a Dirac semimetal and explained it in terms of resonant electron-phonon coupling that occurs when the phonon frequency coincides with the free-carrier cyclotron resonance. In our case, the high magnetic field moves the free-carrier cyclotron frequency completely out of our frequency range, and thus, this mechanism is irrelevant.  The magnetic phonon moment in the present study, instead, comes from phonon anharmonicity and morphic effects due to high magnetic fields, as explained below. Usually, phonons do not present significant magnetic field dependence, but in ionic crystals, the presence of a magnetic field can induce changes in the effective charge, leading to circular dichroism~\cite{ANASTASSAKIS19721091}. For an infrared-active (TO) phonon, the Hamiltonian in the presence of a magnetic field can, in general, be written in the circular basis as 
\begin{equation}
    \mathcal H= \mathcal H_{\text{ph}} + Z_{P S}^{*} E_{P}^{*} Q_{S} + i h_{P S N} E_{P}^{*} Q_{S} H_N,
\end{equation}
where $Z_{P S}^{*}$, $P,S=\{x_R, x_L, z\}$ is the Born effective charge in a circular basis with the $z$ axis (assumed to be the THz probe pulse propagation axis) aligned along the crystallographic $[111]$ direction. Here, $x_R, x_L$ are defined on a plane perpendicular to the $z$ axis. $E_P$ is the applied electric field, $H_N$ is the applied magnetic field, and $Q_S$ represents the phonon mode coordinate in the circular basis. The nonzero terms of the tensor $h_{P S N}$ $P,S,N=\{x_R, x_L, z\}$ are dictated by symmetry constraints. The details of the derivation are presented in the Supplemental Material~\cite{supplemental}. 

For a magnetic field in the $z$ direction, and $E_z = 0$ (parallel to the crystallographic $[111]$ direction), we have 
\begin{equation}
    \mathcal {H} =\mathcal {H}_{\text{ph}} + (Z^{*}+h H_{z} )E_{R} Q_{L}  + (Z^{*}-h H_{z}) E_{L} Q_{R}.
    \label{eq:mcd_ham}
\end{equation}
This result implies that a left-handed electric field will excite a right-handed phonon (composed of a superposition of the degenerate $T_{1u}$ modes) with amplitude $Z^{*}-h H_{z}$. On the other hand, a right-handed electric field will excite a left-handed phonon mode with amplitude $Z^{*}+h H_{z}$. This effect is observed in the data (see Fig.~\ref{fig:magnetic}(a) and (b)) as a difference in the peak heights as a function of magnetic field for LH and RH conductivity. Nevertheless, the model does not contain any explicit temperature dependence for the MCD, as seen in Fig.~\ref{fig:magnetic}(b), where a continuous increase occurs in the studied temperature range. It is possible that the electric field-phonon coupling constant has a temperature dependence, which is beyond our theory, Eq.\,\eqref{eq:mcd_ham}.

In addition to the difference in absorption between left-circularly and right-circularly polarized light, a strong magnetic field can induce changes in the phonon frequency. For phonons with symmetry $T_{1u}$ in the cubic point group $O_h$, group theory indicates that magnetic fields can modify their frequencies to linear~\cite{horton_maradudin_1980} and quadratic order~\cite{ANASTASSAKIS1971563}. The mechanism, as introduced in Ref.~\cite{ANASTASSAKIS1971563} can be summarized as follows. The interactions between the atoms in a crystal lead to a finite energy potential, which determines the frequency of harmonic phonons. The presence of a magnetic field can induce additional symmetry-allowed terms, which, in turn, modify the phonon frequency.
For PbTe, with point group $O_h$ and triple-degenerate $T_{1u}$ phonons, the frequencies are renormalized to
\begin{equation} \label{eq1}
\begin{split}
    &\Omega_{1}(H_z) = \omega_0 - \frac{K_1}{2 \omega_0} H_z+ \frac{K_2}{2 \omega_0} H_z^2, \\
    &\Omega_{2}(H_z) = \omega_0 + \frac{K_1}{2 \omega_0} H_z + \frac{K_2}{2 \omega_0} H_z^2, \\
    &\Omega_{3}(H_z) = \omega_0 + \frac{K_2}{2 \omega_0} H_z^2.
\end{split}
\end{equation}
We relate $\Omega_{1}(H_z)$ and $\Omega_{2}(H_z)$ with the TO phonons. 

In Fig.~\ref{fig:magnetic}(c), the data clearly displays the predicted quadratic magnetic field dependence in the frequencies of the RH and LH TO phonons.  To highlight the Zeeman splitting ($\propto H_z$) and diamagnetic shift ($\propto H^2_z$), we combine the first two equations of \eqref{eq1} to write the energies of the two branches as 
\begin{equation}
    E_\pm(H_z) = E_0 \pm g^* \mu_\mathrm{B} \mu_0H_z + \sigma_\mathrm{dia} \mu_0^2H_z^2,
    \label{eq:fit}
\end{equation}
where $E_0$ is the TO phonon energy at zero magnetic field, $g^*$ is the effective $g$ factor, $\mu_\mathrm{B}$ is the Bohr magneton, $\mu_0$ is the vacuum permeability, and $\sigma_\mathrm{dia}$ is the diamagnetic shift coefficient. The two parameters in Eq.~\eqref{eq:fit}, $g^*$ and $\sigma_\mathrm{dia}$, are related to the coefficients $K_1$ and $K_2$ in Eqs.~\eqref{eq1} as $g^*=K_1\hbar/(2\omega_0\mu_0\mu_B)$ and $\sigma_\mathrm{dia}=K_2\hbar/(2\omega_0\mu_0)$. We used Eq.~\eqref{eq:fit} to fit the data in Fig.~\ref{fig:magnetic}(c), as shown by dashed lines, to obtain $g^* = (4.3 \pm 0.6) \times 10^{-2}$ and $\sigma_\mathrm{dia} = (1.9 \pm 0.2)\times 10^{-4}$\,meV/T$^2$, corresponding to $K_1 = (3\pm0.3)\times10^{-7}$\,THz$^2$/T and $K_2= (2.3\pm0.2)\times10^{-8}$\,THz$^2$/T$^2$. The value of $g^*$ we obtained ($\sim4\times 10^{-2}$) is 3 orders of magnitude larger than the value ($\sim6\times 10^{-5}$) predicted by using density functional theory in Ref.~\cite{JuraschekSpaldin2019PRM} assuming harmonic phonons. This disagreement indicates that anharmonicity is behind the observed large Zeeman splitting.

The diamagnetic shift increases the energies of both Zeeman-split branches with increasing magnetic field. In Fig.~\ref{fig:magnetic}(c), this contribution appears to be relevant for fields above 9\,T when the lower-energy branch reverses its decreasing trend. To the best of our knowledge, this is the first observation of a phononic diamagnetic shift. It has been shown in InSb that the scattering between acoustic phonons can be tuned by a magnetic field control of the anharmonicity, arising from phonon-induced diamagnetism~\cite{jin-diamag}. Such an effect could be extremely important for applications as it indicates that magnetic fields can affect the thermal conductivity. 

In conclusion, we observed circularly polarized phonons in PbTe in magnetic fields. The frequency of these TO phonons exhibited strong anharmonicity in optical conductivity spectra as a function of temperature. Magnetic circular dichroism on the order of 30\% was obtained at 9\,T and below 50\,K. We observed a Zeeman splitting as well as a diamagnetic shift in the phonon frequency as a function of magnetic field up to 25\,T. The magnitude of the Zeeman splitting (linear in magnetic field) can be expressed as a $g$ factor, whose measured value was larger than that previously calculated in the harmonic phonon regime by 3 orders of magnitude, indicating the importance of anharmonicity. We note that a recent theory suggests that a large phonon magnetic moment can arise from the electronic adiabatic response to the ionic circular motion and, in general, can include topological contributions~\cite{ren2021phonon}. On the other hand, the diamagnetic term only becomes important for fields above 9\,T. The results were modeled within group theory in terms of morphic effects and a symmetry-based Hamiltonian. Thus, these results  suggest a novel scheme of controlling soft phonons in an ionic lattice using an external magnetic field.
\nocite{Xu2012,bilbao-sam,QE-2017,QE-2009,doi:10.1063/5.0005082,PhysRevB.88.085117,PhysRevLett.77.3865,PhysRevLett.100.136406,PhysRevB.79.214306,Klemens1966PR,ANASTASSAKIS19721091,horton_maradudin_1980,ANASTASSAKIS1971563,NoeetAl2016OE,MakiharaEtAl2021NC,ChengetAl2020NL,Baydinetal2021FO}

\textit{Acknowledgments.} 
This research was primarily supported by the National Science Foundation through the Center for Dynamics and Control of Materials: an NSF MRSEC under Cooperative Agreement No. DMR-1720595. F.G.G.H. acknowledges financial support from the Brasil@Rice Collaborative Grant, the São Paulo Research Foundation (FAPESP) Grants No. 2015/16191-5 and No. 2018/06142-5, and Grant No. 307737/2020-9 of the National Council for Scientific and Technological Development (CNPq). M. R-V. was supported by the LANL LDRD Program and by the U.S. Department of Energy, Office of Science, Basic Energy Sciences, Materials Sciences and Engineering Division, Condensed Matter Theory Program. G. A. F. acknowledges additional support from Grants No. NSF DMR-1949701 and No. NSF DMR-2114825. J. T. and I. K. acknowledge the support from the Japan Society for the Promotion of Science (JSPS) (KAKENHI Grant No. 20H05662). 

\bibliography{bib,bib2}

\end{document}